\documentclass[prd,reprint,nofootinbib,superscriptaddress,preprintnumbers,showpacs]{revtex4-1}
\usepackage{amsmath}	
\usepackage{amssymb}	
\usepackage{amstext}	
\usepackage{latexsym}	
\usepackage{bm}			
\usepackage{graphicx,color}	
\usepackage{slashed}	
\usepackage{makecell}	
\usepackage{multirow}	
\usepackage{caption}	
\usepackage{hyperref}	

\allowdisplaybreaks[4]
\pdfoutput=1

\begin{document}
\title{The Theoretical Study of $p\bar{p}\to\bar{\Lambda}\Sigma\eta$ Reaction}
\date{\today}
\author{Aojia Xu}
\affiliation{School of Physics, Dalian University of Technology, Dalian 116024, People's Republic of China}
\author{Ruitian Li}
\affiliation{School of Physics, Dalian University of Technology, Dalian 116024, People's Republic of China}
\author{Xuan Luo}
\affiliation{School of Physics and Optoelectronics Engineering, Anhui University, Hefei 230601, People's Republic of China}
\author{Hao Sun}
\email{haosun@dlut.edu.cn}
\affiliation{School of Physics, Dalian University of Technology, Dalian 116024, People's Republic of China}
\begin{abstract}
We study the production of hyperon resonances in the $p\bar{p}\to\bar{\Lambda}\Sigma\eta$ reaction within an effective Lagrangian approach. The model includes the production of $\Sigma(1750)$ and $\Lambda(1670)$ in the intermediate state excited by the $K$ and $K^*$ meson exchanges between the initial proton and antiproton. Due to the large coupling of $\Sigma(1750)\Sigma\eta$ vertex, $\Sigma(1750)$ is found a significant contribution near the threshold in this reaction. We provide total and differential cross section predictions for the reation and discuss the possible influence of $\Sigma(1750)\Sigma\eta$ vertex coupling and model parameters, which will be useful in future experimental studies. This reaction can provide a platform for studying the features of $\Sigma(1750)$ resonance, especially the coupling to $\Sigma\eta$ channel.
\end{abstract}
\maketitle
\section{introduction}
The investigation of the meson-baryon interactions at low energies plays an important role in exploring the features of hyperon resonances. However, experiments on hyperon resonances are not as extensive as those on nucleon resonances. Most of our current knowledge about $\Sigma$ hyperon resonances has come from the analysis of experimental data in the $\Lambda\pi$ and $\bar{K}N$ channels~\cite{Prakhov:2008dc, Cameron:1980nv, Morris:1978ia, Ponte:1975bt, Rutherford-London:1975zvn, Mast:1975pv, Jones:1974at, Baxter:1973ggf, Armenteros:1970eg}. In addition to $\bar{K}N$ scattering reactions, others such as LEPS~\cite{LEPS:2016ljn, LEPS:2009isz, Niiyama:2008rt}, CLAS~\cite{CLAS:2021osv, CLAS:2013rxx}, COSY~\cite{Zychor:2008ct, Zychor:2005sj} have attempted to further generate excited hyperon resonances from $\gamma N$ and $NN$ collisions.

Because of the isospin conservation, the $\Sigma\eta$ channel has a special significance for which it is a pure $I=1$ channel that only coupled to $\Sigma$ hyperon resonances. However, even with this advantage, the researches on $\Sigma$ hyperon resonances are still relatively few. Up to now, only one $\Sigma$ hyperon resonance, $\Sigma(1750)$, was found to be well coupled to the $\Sigma\eta$ channel in the Particle Data Group (PDG)~\cite{Workman:2022ynf} book. While the decay branching ratios of other $\Sigma$ hyperon resonances to this channel are still not well identified, it is possible that other resonances do have rather weak coupling to the $\Sigma\eta$ channel, thus making it difficult to study their coupling to $\Sigma\eta$. It is also the large coupling that makes it possible to distinguish $\Sigma(1750)$ from other $\Sigma$ hyperon resonances in $\Sigma\eta$ channel. Moreover, the threshold energy of $\Sigma\eta$ channel is about 1.74~GeV, which is very close to the mass of $\Sigma(1750)$, providing a suitable place to investigate the features of $\Sigma(1750)$ resonance.

Nevertheless, the coupling of $\Sigma(1750)$ and $\Sigma\eta$ has rarely been studied in previous researches. In the current particle collision experiments, the coupling of $\Sigma(1750)$ and $\Sigma\eta$ channel has only been found in $K^-p\to\Sigma^0\eta$ reactions~\cite{Jones:1974si}. A chiral $\bar{K}N$ interaction model was used in Refs.~\cite{Feijoo:2022zfn, Feijoo:2021zau} to fit the experimental data of the production cross section and analyze the possible resonances in the reaction process. Ref.~\cite{Nogueira-Santos:2023usb} used a effective chiral Lagrangian method to  study $\eta-$baryon interactions at low energies in the $\eta B\to\eta B$ process, including the coupling of $\Sigma(1750)$ and $\Sigma\eta$. Some works have investigated the partial wave analysis of $\bar{K}N$ scattering, such as Ref.~\cite{Zhang:2013sva} using a global multichannel fit for all the $\bar{K}N$ scattering reactions; Refs.~\cite{Kamano:2015hxa, Kamano:2014zba} used a dynamic coupled channel model to establish the spectrum of $\Sigma$ hyperon resonances and extract the resonance parameters, however, the $\bar{K}N\to\Sigma\eta$ reaction was not taken into account. In recent years, $\bar{\mathrm{P}}$ANDA collaboration has accumulated a lot of experimental datas in $p\bar{p}$ scattering\cite{Rieger:2023vyd, PANDA:2023ljx, PANDA:2022frd, Nerling:2021bxo, PANDA:2021ozp, PANDA:2020hmi}, we hope that the reaction we proposed will be helpful to search for $\Sigma(1750)$ resonance in future experiments. Furthermore, the high-intensity heavy-ion accelerator facility (HIAF)\cite{Zhou:2022pxl} in China will be put into use in the near future, which is very suitable for exploring the feature of hyperon resonance. Such experiment will definitely offer valuable data for improving our knowledge of the strong interaction and hyperon spectroscopy.

In the present work, we propose the $p\bar{p}\to\bar{\Lambda}\Sigma\eta$ reaction that can be used to investigate the features of $\Sigma(1750)$ resonance. We investigate the reaction by using an effective Lagrangian approach, focusing on the production of $\Sigma(1750)$ hyperon resonance. The approach of effective Lagrangian calculating the reaction cross section is widely used to investigate the process of particle collisions for exploring the reaction mechanism between initial and final particles~\cite{Shi:2023xfz, Kim:2021wov, Liu:2020wlg, Wang:2017sxq, Xie:2014zga, Gao:2013qta, Gao:2012zh, Sharov:2011xq, Man:2011np, Oh:2006hm}. Near the threshold of $\Sigma\eta$ channel, only $\Sigma(1750)$ was found to have a relatively large decay branch ratio to $\Sigma\eta$ channel, which can be naturally regarded that $\Sigma(1750)$ has a large coupling to $\Sigma\eta$ channel. Besides, the resonance contribution in the $\Lambda\eta$ channel should also be taken into account. Same as $\Sigma(1750)$, here we only need to consider the contribution of $\Lambda(1670)$. In our model, the $\Sigma(1750)$ and $\Lambda(1670)$ resonances are excited by the $K$ and $K^*$ meson exchanges between the initial proton and antiproton. Other meson exchanges are forbidden by the law of isospin conservation. The predictions of the total cross section and angular distribution, as well as invariant mass distribution are presented in our work, which will be helpful for future comparison with the experimental data. We also provide a discussion for the dependence of total and differential cross sections on model parameters.

Our work is organized as follows. In Sec.~II, we introduce the formalism and ingredients necessary of each amplitude in our model and obtain the concrete form of amplitudes. The numerical results of the total and differential cross sections for $p\bar{p}\to\bar{\Lambda}\Sigma\eta$ reaction are presented in Sec.~III. Finally, a short conclusion is made in Sec.~IV.

\section{Formalism}
\begin{figure}[htpb]
	\centering
	\includegraphics[width=0.35\textwidth]{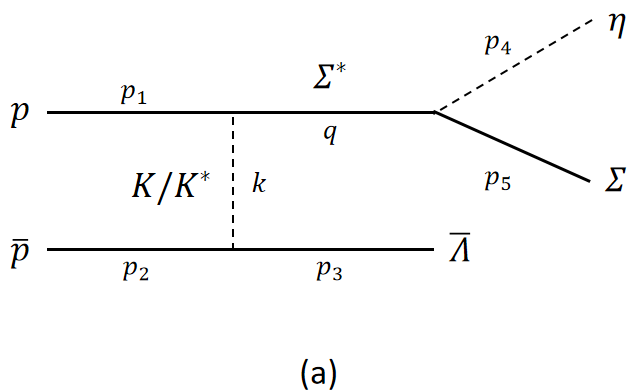}\hypertarget{1a}{}\\
	\includegraphics[width=0.35\textwidth]{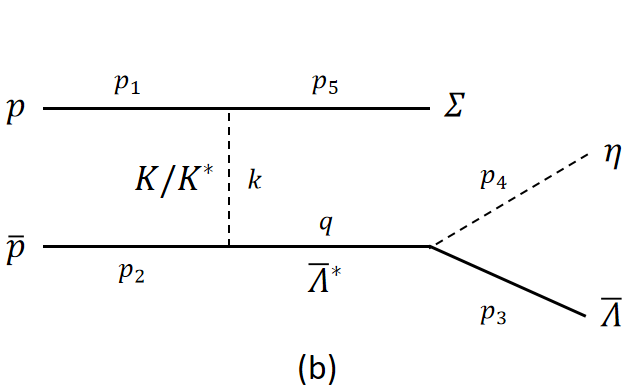}\hypertarget{1b}{}
	\captionsetup{justification=raggedright}
	\caption{(a) u- (b) t-channel exchanges Feynman diagrams for $p\bar{p}\to\bar{\Lambda}\Sigma\eta$ reaction.}
	\label{1}
\end{figure}

Within our approach, the production mechanism of the $\Sigma(1750)$ and $\Lambda(1670)$ resonances in the reaction $p\bar{p}\to\bar{\Lambda}\Sigma\eta$ consists of the standard $t-$ and $u-$channel as shown in Fig.~\ref{1}. In view of $\Sigma(1750)$ has a relatively large coupling to the $\Sigma\eta$ channel, we expect it may give a significant contribution in the reaction. Because of charge, the $K$ and $K^*$ exchanges are present only for the charged $K^+$ and $K^{*+}$. 

The production amplitude is calculated from the following effective Lagrangians,
\begin{equation}
	\begin{split}
		\mathcal{L}_{\Lambda KN}=&-\dfrac{g_{\Lambda KN}}{m_\Lambda+m_N}\bar{\Lambda}\gamma_5\slashed{\partial}\bar{K}N+h.c.,\\
		\mathcal{L}_{\Sigma^*KN}=&\sqrt{2}ig_{\Sigma^*KN}\bar{\Sigma}^*KN+h.c.,\\
		\mathcal{L}_{\Sigma\eta\Sigma^*}=&-ig_{\Sigma\eta\Sigma^*}\bar{\Sigma}\eta\Sigma^*+h.c.,\\
		\mathcal{L}_{\Sigma KN}=&-\dfrac{\sqrt{2}g_{\Sigma KN}}{m_\Sigma+m_N}\bar{\Sigma}\gamma_5\slashed{\partial}\bar{K}N+h.c.,\\
		\mathcal{L}_{\Lambda^*KN}=&ig_{\Lambda^*KN}\bar{\Lambda}^*KN+h.c.,\\
		\mathcal{L}_{\Lambda\eta\Lambda^*}=&-ig_{\Lambda\eta\Lambda^*}\bar{\Lambda}\eta\Lambda^*+h.c.,
	\end{split}
\end{equation}
and for $K^*$ exchange,
\begin{equation}
	\begin{split}
		\mathcal{L}_{\Lambda K^*N}=&-g_{\Lambda K^*N}\bar{\Lambda}\left(\gamma^\mu+\dfrac{\kappa_{\Lambda K^*N}}{2m_N}(p_{K^*}^\mu-\slashed{p}_{K^*}\gamma^\mu)\right)K^*_\mu N\\
		&+h.c.,\\
		\mathcal{L}_{\Sigma^*K^*N}=&ig_{\Sigma^*K^*N}\bar{\Sigma}^*\gamma_5\gamma^\mu K^*_\mu N+h.c.,\\
		\mathcal{L}_{\Sigma K^*N}=&-g_{\Sigma K^*N}\bar{\Sigma}\left(\gamma^\mu+\dfrac{\kappa_{\Sigma K^*N}}{2m_N}(p_{K^*}^\mu-\slashed{p}_{K^*}\gamma^\mu)\right)K^*_\mu N\\
		&+h.c.,\\
		\mathcal{L}_{\Lambda^*K^*N}=&ig_{\Lambda^*K^*N}\bar{\Lambda}^*\gamma_5\gamma^\mu K^*_\mu N+h.c.,
	\end{split}
\end{equation}
where $\kappa_{\Lambda K^*N}=2.76$ and $\kappa_{\Sigma K^*N}=-2.33$~\cite{Wang:2017tpe} are the anomalous magnetic moments. The coupling constants $g_{\Lambda KN}$, $g_{\Sigma KN}$, $g_{\Lambda K^*N}$ and $g_{\Sigma K^*N}$ can be determined by the SU(3) predictions~\cite{Ronchen:2012eg}, which give the values that $g_{\Lambda KN}=-13.99~\mathrm{GeV}^{-1}$, $g_{\Sigma KN}=2.69~\mathrm{GeV}^{-1}$, $g_{\Lambda K^*N}=-6.21~\mathrm{GeV}^{-1}$ and $g_{\Sigma K^*N}=-4.25~\mathrm{GeV}^{-1}$. And we take the value for $\Lambda^*K^*N$ coupling from Ref.~\cite{Xiao:2015zja}. For $\Sigma^*K^*N$ coupling, we adopt the same value as $\Sigma^*KN$, approximatively. Other constants are determined from the partial decay widths, given in Table.~\hyperlink{tab1}{1}. It should be noted that we use the average values of the branching ratios listed in PDG~\cite{Workman:2022ynf}. Due to the masses of $\Sigma(1750)$ and $\Lambda(1670)$ are very close to $\Sigma\eta$ and $\Lambda\eta$ thresholds, respectively, taking their finite widths into account is essential. We include the finite width effect by using the following formula as~\cite{Wang:2023lnb, Roca:2005nm}
\begin{equation}
	\begin{split}
		\Gamma_{\Sigma^*\to\Sigma\eta}=&-\dfrac{1}{\pi}\int_{(m_{\Sigma^*}-2\Gamma_{\Sigma^*})^2}^{(m_{\Sigma^*}+2\Gamma_{\Sigma^*})^2}ds\Gamma_{\Sigma^*\to\Sigma\eta}(\sqrt{s})\\
		&\times\Theta(\sqrt{s}-m_\Sigma-m_\eta)\mathrm{Im}\left\{\dfrac{1}{s-m_{\Sigma^*}^2+im_{\Sigma^*}\Gamma_{\Sigma^*}}\right\}.
	\end{split}
\end{equation}
\begin{table}[htbp]
	\renewcommand{\arraystretch}{1.8}
	\tabcolsep=1.6mm
	\captionsetup{justification=raggedright}
	\caption*{TABLE~I. Coupling constants used in this work.}\hypertarget{tab1}{}
	\begin{tabular}[b]{ccccc}
		State & \makecell{Width\\(MeV)} & \makecell{Decay\\channel} & \makecell{Branching ratio\\adopted} & $g^2/4\pi$\\
		$\Sigma(1750)$ & 206 & $\Sigma\eta$ & 0.35  & $4.11\times10^{-1}$\\
		 &  & $NK$ & 0.09  & $1.66\times10^{-2}$\\
		$\Lambda(1670)$ & 32 & $\Lambda\eta$ & 0.175  & $6.06\times10^{-2}$\\
		 &  & $NK$ & 0.25  & $0.82\times10^{-2}$
	\end{tabular}
\end{table}

Since hadrons are not pointlike particles, it is necessary to consider a form factor at each vertex, which can parameterize the structure of the hadron. Here, we introduce the form factor for intermediate baryons as
\begin{equation}
	f_B(q_B^2)=\dfrac{\Lambda_B^4}{\Lambda_B^4+(q_B^2-m_B^2)^2},
\end{equation}
with $q_B$ and $m_B$ the four-momentum and mass of intermediate hadron, respectively. The cut-off parameter for $\Lambda^*$ exchange is taken as $\Lambda_{\Lambda^*}=1.5$~GeV.

For $K$ meson and $K^*$ meson exchange diagrams, we introduce the form factor as
\begin{equation}
	f_M(k_M^2)=\left(\dfrac{\Lambda_M^2-m_M^2}{\Lambda_M^2-k_M^2}\right)^n,
\end{equation}
where $k_M$ and $m_M$ denote the four-momentum and mass of exchanged meson, respectively. Here, we take $\Lambda_K=1.1$~GeV~\cite{Oh:2006hm} and $\Lambda_{K^*}=1.5$~GeV~\cite{Liu:2011sw, Liu:2012ge} for the corresponding meson exchange. In the caculation, $n=1$ for $K$ exchange and $n=2$ for $K^*$ exchange~\cite{Huang:2012xj} are adopted.

The propagators for the exchanged particles are expressed as
\begin{equation}
	G_K(k)=\dfrac{i}{(k^2-m_K^2)},
\end{equation}
for $K$ meson,
\begin{equation}
	G_{K^*}^{\mu\nu}(k)=-i\left(\dfrac{g^{\mu\nu}-\frac{k^\mu k^\nu}{m_{K^*}^2}}{k^2-m_{K^*}^2}\right),
\end{equation}
for $K^*$ meson, and
\begin{equation}
	G^\frac{1}{2}(q)=\dfrac{i(\slashed{q}\pm M_B)}{q^2-M_B^2+iM_B\Gamma_B},
\end{equation}
for spin-1/2 baryons with '+' and '-' correspond to particle and antiparticle respectively, where $k$ and $q$ are the four-momentum; $M_B$ and $\Gamma_B$ are the mass and width of intermediate baryons.

With the ingredients presented above, the total scattering amplitudes of $p\bar{p}\to\bar{\Lambda}\Sigma\eta$ reaction can be written as
\begin{equation}
	\begin{split}
		\mathcal{M}_a^K=&\dfrac{\sqrt{2}ig_{\Sigma\eta\Sigma^*}g_{\Sigma^*Kp}g_{\Lambda Kp}}{m_\Lambda+m_p}f_K^2(k_1)f_{\Sigma^*}(q_1)\bar{u}(p_5,s_5)\\
		&\times G_{\Sigma^*}(q_1)u(p_1,s_1)G_K(k_1)\bar{v}(p_2,s_2)\gamma_5\slashed{k}_1v(p_3,s_3),\\
		\mathcal{M}_b^K=&\dfrac{\sqrt{2}ig_{\Lambda\eta\Lambda^*}g_{\Lambda^*Kp}g_{\Sigma Kp}}{m_\Sigma+m_p}f_K^2(k_2)f_{\Lambda^*}(q_2)\bar{v}(p_2,s_2)\\
		&\times G_{\Lambda^*}(q_2)v(p_3,s_3)G_K(k_2)\bar{u}(p_5,s_5)\gamma_5\slashed{k}_2u(p_1,s_1),\\
		\mathcal{M}_a^{K^*}=&-g_{\Sigma\eta\Sigma^*}g_{\Sigma^*K^*p}g_{\Lambda K^*p}f_{K^*}^2(k_1)f_{\Sigma^*}(q_1)\bar{u}(p_5,s_5)\\
		&\times G_{\Sigma^*}(q_1)\gamma_5\gamma^\mu u(p_1,s_1)G_{K^*\mu\nu}(k_1)\bar{v}(p_2,s_2)\\
		&\times\left(\gamma^\nu+\dfrac{\kappa_{\Lambda K^*p}}{2m_p}(k_1^\nu-\slashed{k}_1\gamma^\nu)\right)v(p_3,s_3),\\
		\mathcal{M}_b^{K^*}=&-g_{\Lambda\eta\Lambda^*}g_{\Lambda^*K^*p}g_{\Sigma K^*p}f_{K^*}^2(k_2)f_{\Sigma^*}(q_2)\bar{v}(p_2,s_2)\\ 
		&\times\gamma^\mu\gamma_5G_{\Lambda^*}(q_2)v(p_3,s_3)G_{K^*\mu\nu}(k_2)\bar{u}(p_5,s_5)\\
		&\times\left(\gamma^\nu+\dfrac{\kappa_{\Sigma K^*p}}{2m_p}(k_2^\nu-\slashed{k}_2\gamma^\nu)\right)u(p_1,s_1).
	\end{split}
\end{equation}
The $p_1$, $p_2$, $p_3$ and $p_5$ represent the four-momentums of the $p$, $\bar{p}$, $\bar{\Lambda}$ and $\Sigma$ baryon, respectively. $k_1$ and $k_2$ correspond to the four-momentum of exchanged meson in Fig.~\hyperlink{1a}{1(a)} and Fig.~\hyperlink{1b}{1(b)}, respectively. $q_1$ and $q_2$ has the same meaning as $k_1$, $k_2$, but for $\Sigma(1750)$ and $\bar{\Lambda}(1670)$.

The differential and total cross sections for this reaction can be obtained through
\begin{equation}
	\begin{split}
		d\sigma=&\dfrac{(2\pi)^4}{4\sqrt{(p_1\cdot p_2)^2-m_p^4}}\left(\dfrac{1}{4}\sum|\mathcal{M}|^2\right)d\Phi_3\\
		=&\dfrac{1}{(2\pi)^4}\dfrac{1}{\sqrt{(p_1\cdot p_2)^2-m_p^4}}\dfrac{|\vec{p}_3||\vec{p}_5^*|}{32\sqrt{s}}\left(\dfrac{1}{4}\sum|\mathcal{M}|^2\right)\\
		&dm_{\Sigma\eta}d\Omega_5^*d\mathrm{cos}\theta_3,
	\end{split}
\end{equation}
where $p_1$, $p_2$ represent the four-momentum of the initial particles $p$, $\bar{p}$ at total center-of-mass frame; $\vec{p}_5^*$ stands for the three-momentum of the $\Sigma$ baryon in the center-of-mass frame of $\Sigma\eta$ pair.

\section{results}
\begin{figure}[htb]
	\centering
	\includegraphics[width=0.45\textwidth]{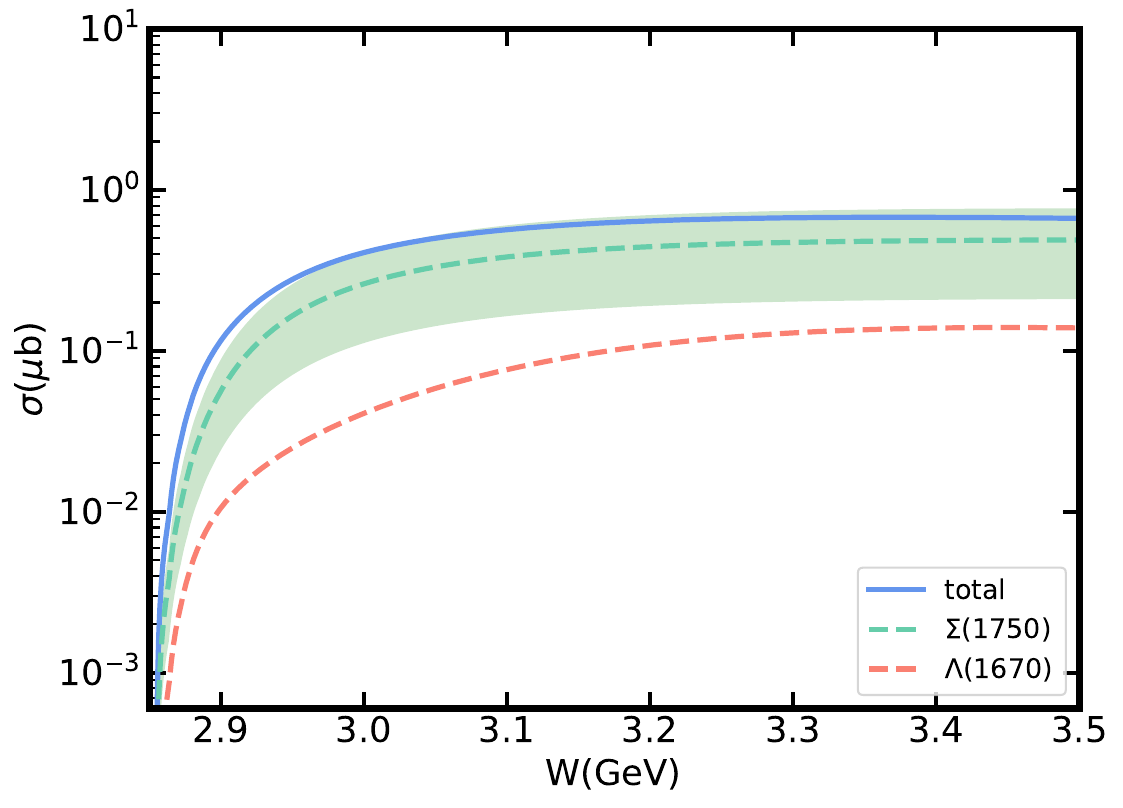}
	\captionsetup{justification=raggedright}
	\caption{Total cross section vs center of mass energy W for $p\bar{p}\to\bar{\Lambda}\Sigma\eta$ reaction. The blue curve represents the total cross section including all the contributions in Fig.~\ref{1}. The green dashed and red dashed curves are the contributions of $\Sigma(1750)$ and $\Lambda(1670)$  respectively, with the branching ratio of $\Sigma^*\to\Sigma\eta$ takes the middle value $35\%$. The green band is the branching ratio of $\Sigma^*\to\Sigma\eta$ from $15\%$ to $55\%$.}
	\label{2}
\end{figure}
In this section, we will present the theoretical results of the $p\bar{p}\to\bar{\Lambda}\Sigma\eta$ reaction calculated by the model in the previous section, including the total cross section and the differential cross section. Firstly, we consider the effects of the branching ratio $Br(\Sigma^*\to\Sigma\eta)$ on the total cross section by fixing the cut-off parameters $\Lambda_{\Sigma^*}=\Lambda_{\Lambda^*}=1.5$~GeV. In Fig.~\ref{2}, we plot the total cross section from the reaction threshold up to 3.5~GeV, together with the individual contributions of $\Sigma(1750)$ and $\Lambda(1670)$ resonances. Both contributions of $K$ and $K^*$ exchanges are taken into account. It is obvious that $\Sigma(1750)$ plays a dominant role of this reaction. Even if we take the minimum value of branching ratio that $Br(\Sigma^*\to\Sigma\eta)=15\%$, the contribution of $\Sigma(1750)$ is significantly larger than that of $\Lambda(1670)$ in this reaction. The significant contribution of $\Sigma(1750)$ is due in part to the coupling of $\Sigma(1750)\Sigma\eta$ vertex is strong compared to that of $\Lambda(1670)\Lambda\eta$. But more importantly, the coupling of $\Lambda KN$ vertex is more than 5 times that of $\Sigma KN$ vertex. Since $\Sigma(1750)$ plays a dominant role near the threshold, it can be considered that this reaction provides a good place for studying the nature of $\Sigma(1750)$ resonance.

\begin{figure}[htbp]
	\centering
	\includegraphics[width=0.45\textwidth]{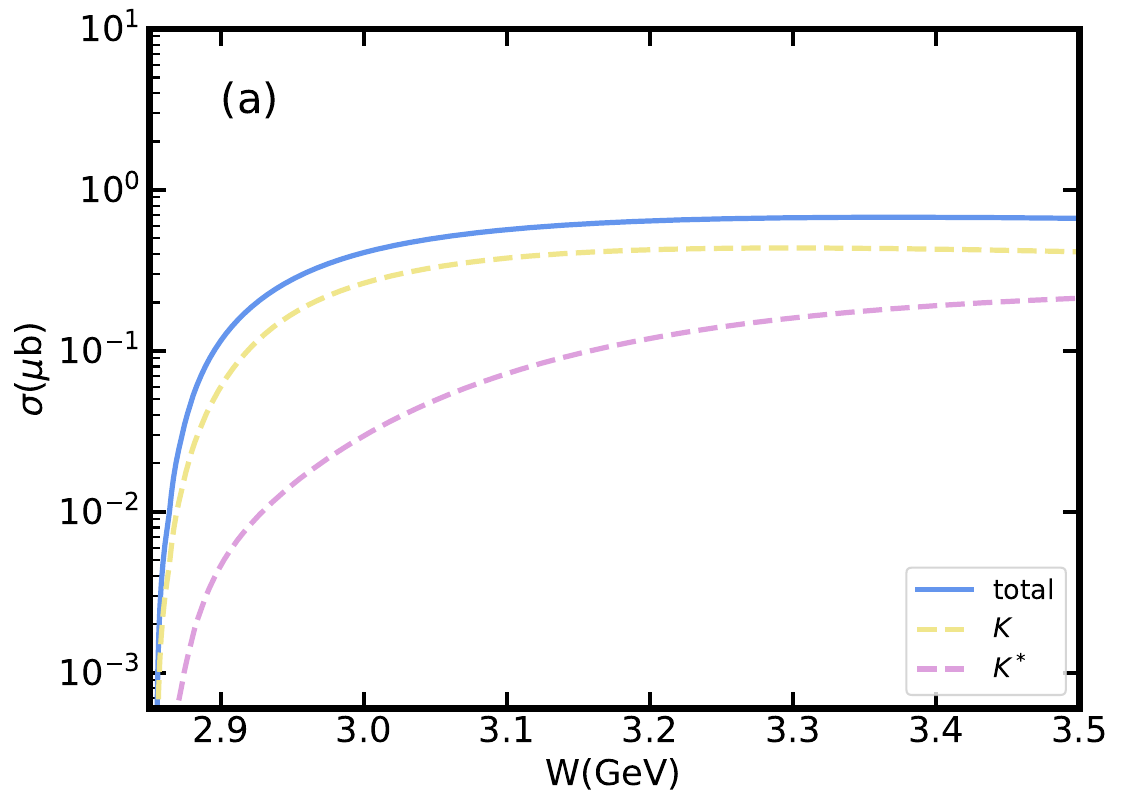}
	\includegraphics[width=0.45\textwidth]{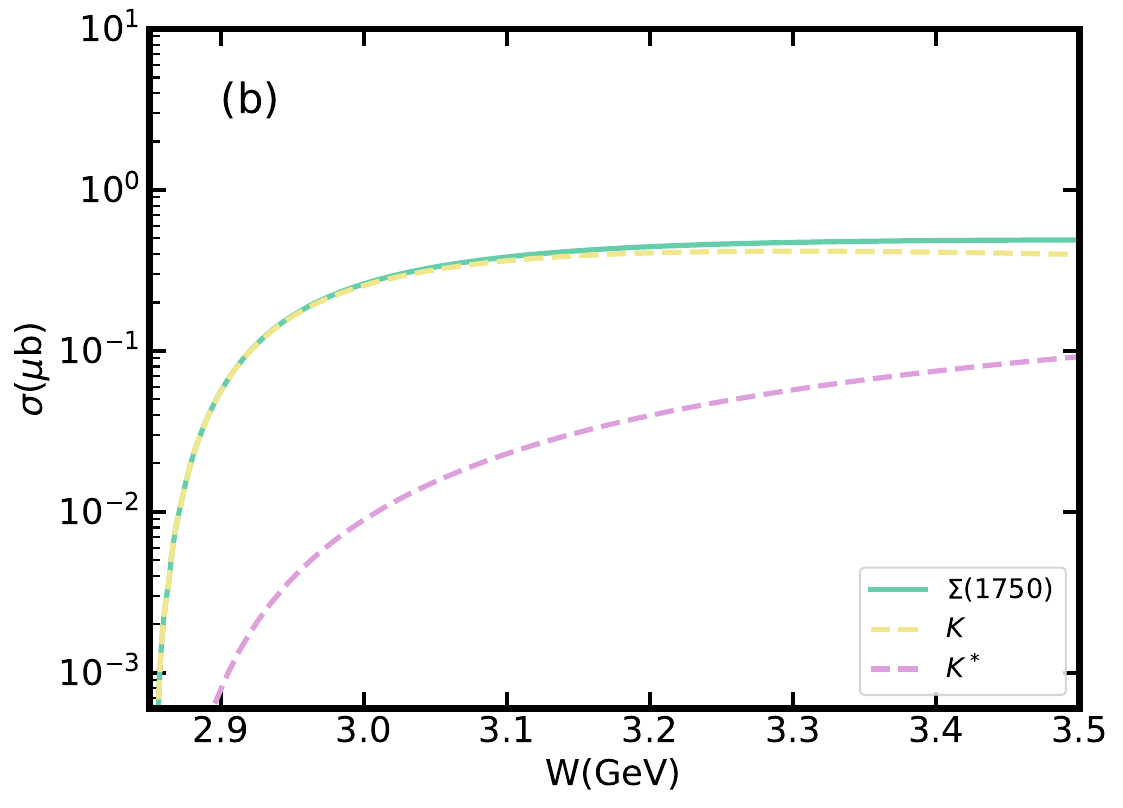}
	\includegraphics[width=0.45\textwidth]{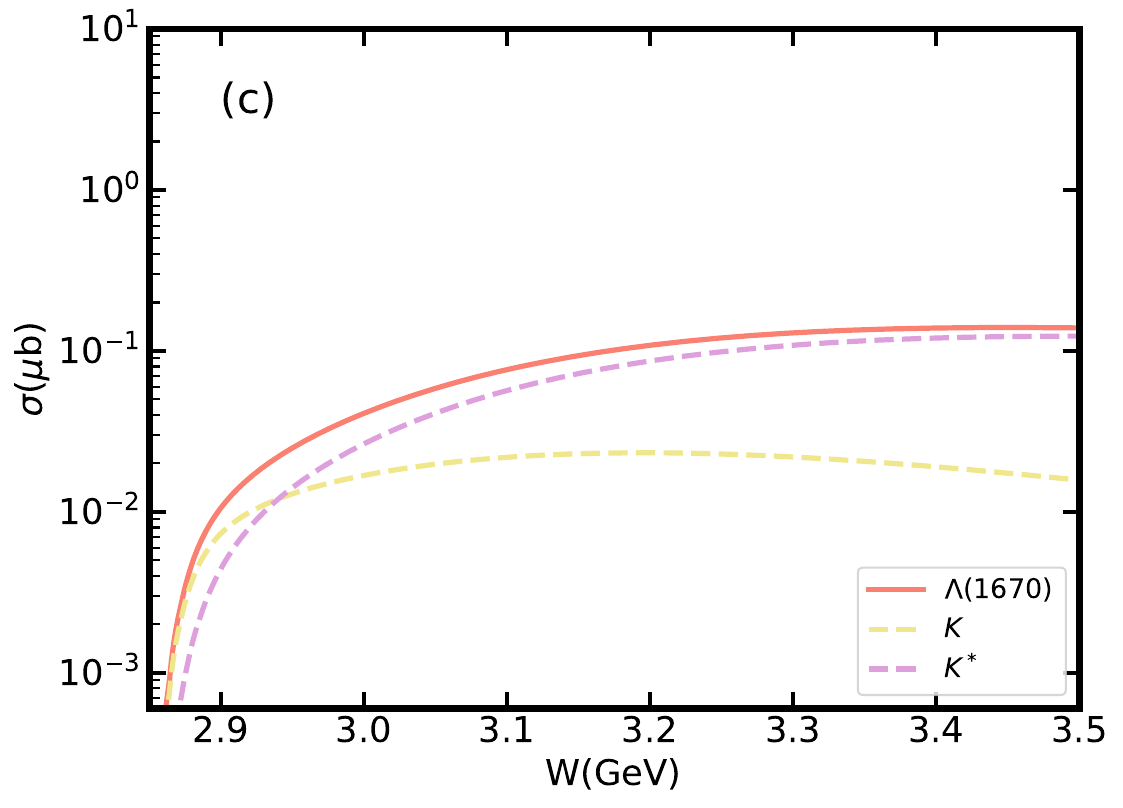}
	\hypertarget{3c}{}
	\captionsetup{justification=raggedright}
	\caption{The cross sections vs center of mass energy W from $K^+$ and $K^{*+}$ exchanges. The blue, green and red curves are the total cross section, $\Sigma(1750)$ and $\Lambda(1670)$ contributions, respectively. The yellow dashed and purple dashed represent the $K^+$ and $K^{*+}$ mesons exchanged contributions.}
	\label{3}
\end{figure}
Because of the uncertainty of the form factor, it is necessary to consider the effect of form factor on the cross section. For this reaction, only the effect of form factor on cross sections of $K^+$ and $K^{*+}$ exchanges needs to be considered. In Fig.~\ref{3}, we show the cross sections from $K^+$ and $K^{*+}$ exchanges compare to the total cross section and $\Sigma(1750)$, $\Lambda(1670)$ resonances. The results show clearly that $K^+$ exchange gives the dominant contribution in total cross section and $\Sigma(1750)$ contribution. The dominant role of $K^+$ exchange in $\Sigma(1750)$ can be attributed to the relatively large $\Lambda Kp$ coupling. While in Fig.~\hyperlink{3c}{3(c)}, it shows that $K^+$ exchange is the main contribution of $\Lambda(1670)$ near the threshold. With the energy of the center of mass increasing, the contribution of $K^{*+}$ exchange is becoming more and more significant. Moreover, due to the influence of $K^{*+}$ exchange, the cross section of $\Lambda(1670)$ shows a relatively obvious trend of gradual increase.

\begin{figure}[b]
	\centering
	\includegraphics[width=0.45\textwidth]{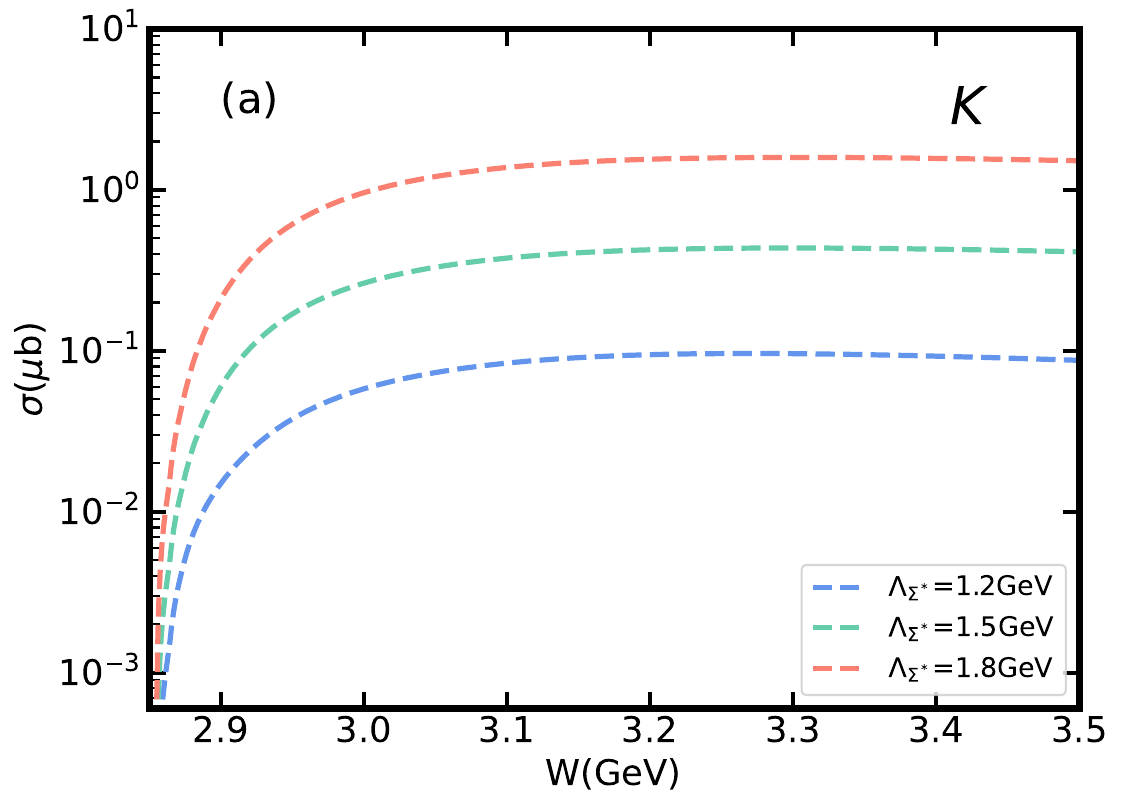}
	\includegraphics[width=0.45\textwidth]{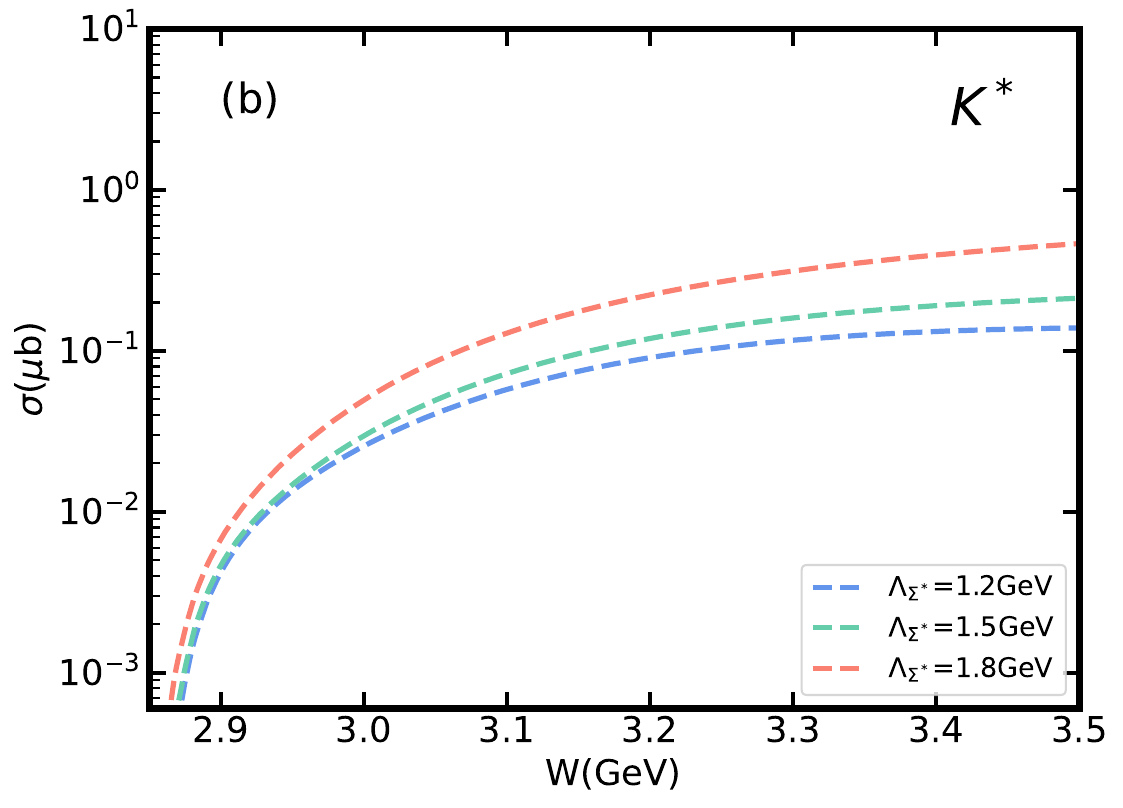}
	\captionsetup{justification=raggedright}
	\caption{The cross sections vs center of mass energy W from the $K^+$ and $K^{*+}$ mesons exchanged contributions with the different values of cut-off parameter $\Lambda_{\Sigma^*}$.}
	\label{4}
\end{figure}

\begin{figure}[t]
	\centering
	\includegraphics[width=0.45\textwidth]{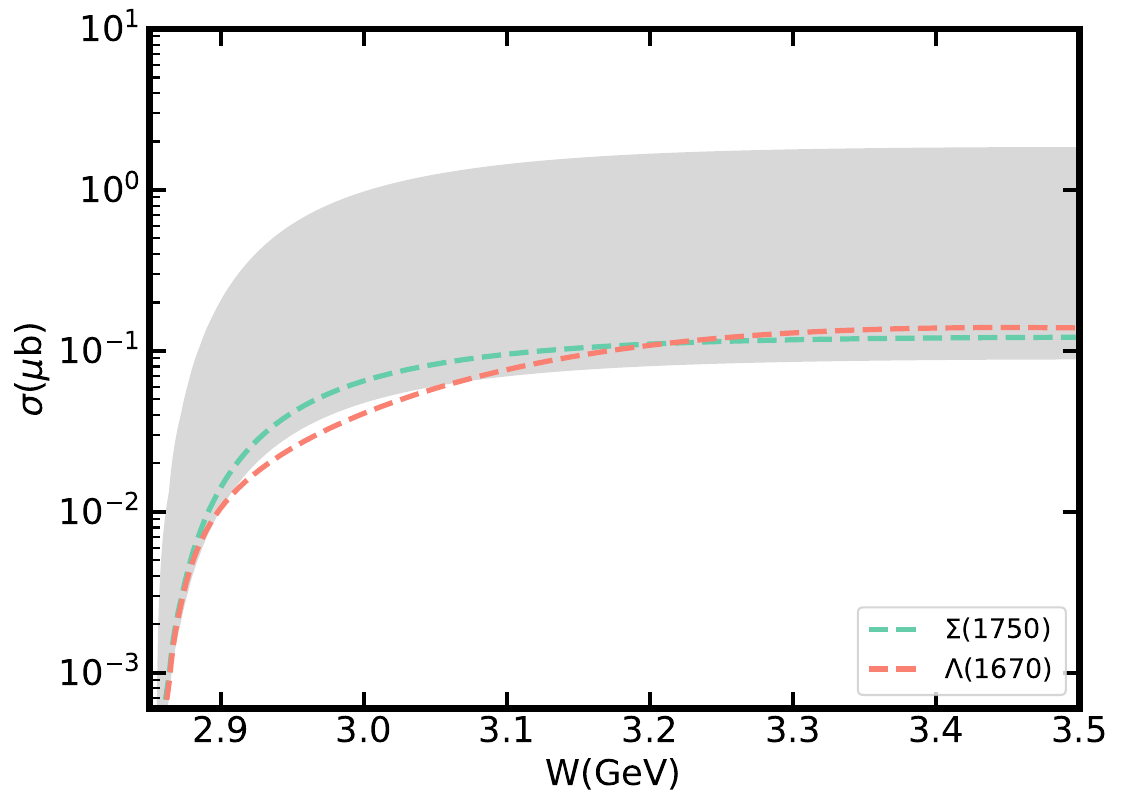}
	\captionsetup{justification=raggedright}
	\caption{The cross sections vs center of mass energy W from the individual contributions of $\Sigma(1750)$ and $\Lambda(1670)$. The green dashed curve represents the contribution of $\Sigma(1750)$ with $\Lambda_{\Sigma^*}=1.25$~GeV.The band corresponds to the results of $\Sigma(1750)$ by varying the cut-off parameter $\Lambda_{\Sigma^*}$ for $\Sigma(1750)KN$ vertex from $1.2$ to $1.8$~GeV.}
	\label{5}
\end{figure}

Next, we consider the dependence of the cross section on the model parameter $\Lambda_{\Sigma^*}$ introduced by the form factor. In Fig.~\ref{4}, we present the cross sections of $K^+$ and $K^{*+}$ exchanges with three cut-off parameters $\Lambda_{\Sigma^*}=1.2, 1.5, 1.8$~GeV. Whether near the threshold or at higher energies, the $K^+$ exchange is more sensitive to the change of cut-off parameter $\Lambda_{\Sigma^*}$. In addition, due to the dominant role of the $K^+$ exchange in $\Sigma(1750)$ contribution, the $\Sigma(1750)$ contribution also has a strong dependence on the value of $\Lambda_{\Sigma^*}$, as shown in Fig.~\ref{5}. When $\Lambda_{\Sigma^*}$ is larger than 1.25~GeV, $\Sigma(1750)$ plays a main role in total cross section. If we take $Br(\Sigma^*\to\Sigma\eta)=15\%$, this value will be raised to 1.4~GeV. At $\Lambda_{\Sigma^*}=1.25$~GeV, $\Sigma(1750)$ dominates at lower energy, and $\Lambda(1670)$ gradually contributes more than $\Sigma(1750)$ as the energy increases. When $\Lambda_{\Sigma^*}$ is below 1.25~GeV, the $\Lambda(1670)$ becomes the dominant contribution. However, since the cut-off parameter is often regarded as a free parameter, more experimental data are needed to determine it.

\begin{figure*}[htbp]
	\centering
	\includegraphics[width=0.3\textwidth]{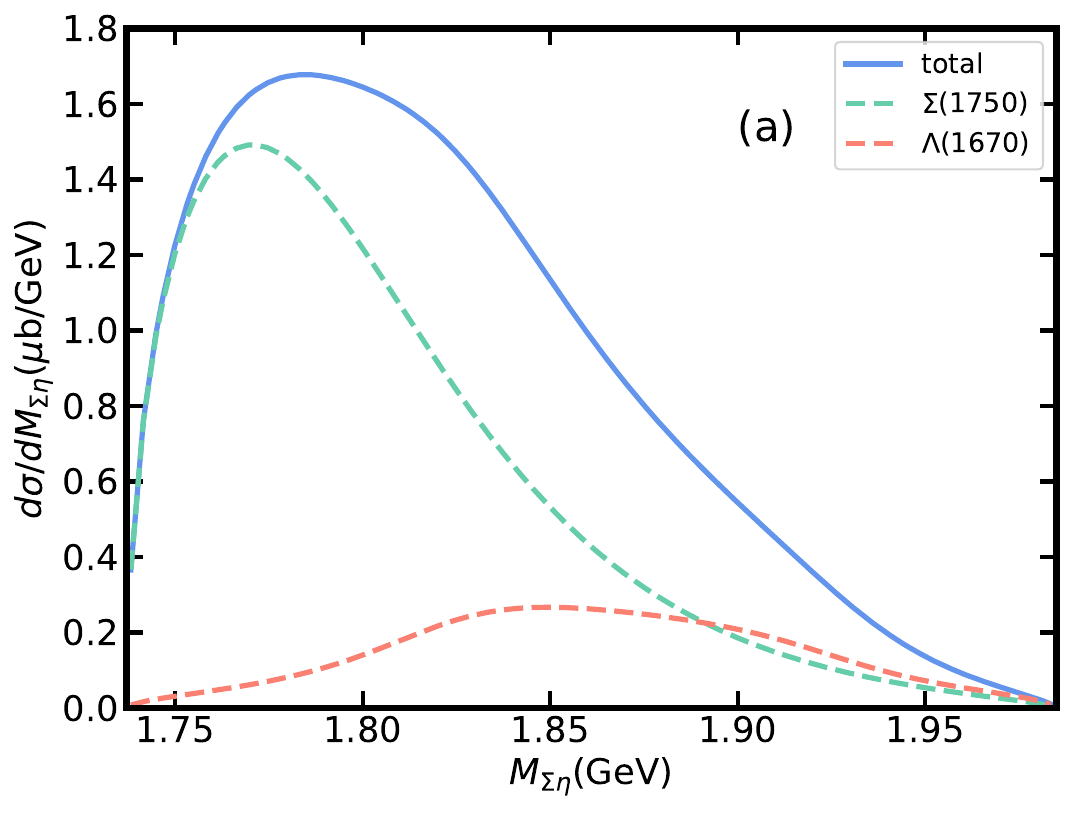}\hypertarget{6a}{}
	\includegraphics[width=0.3\textwidth]{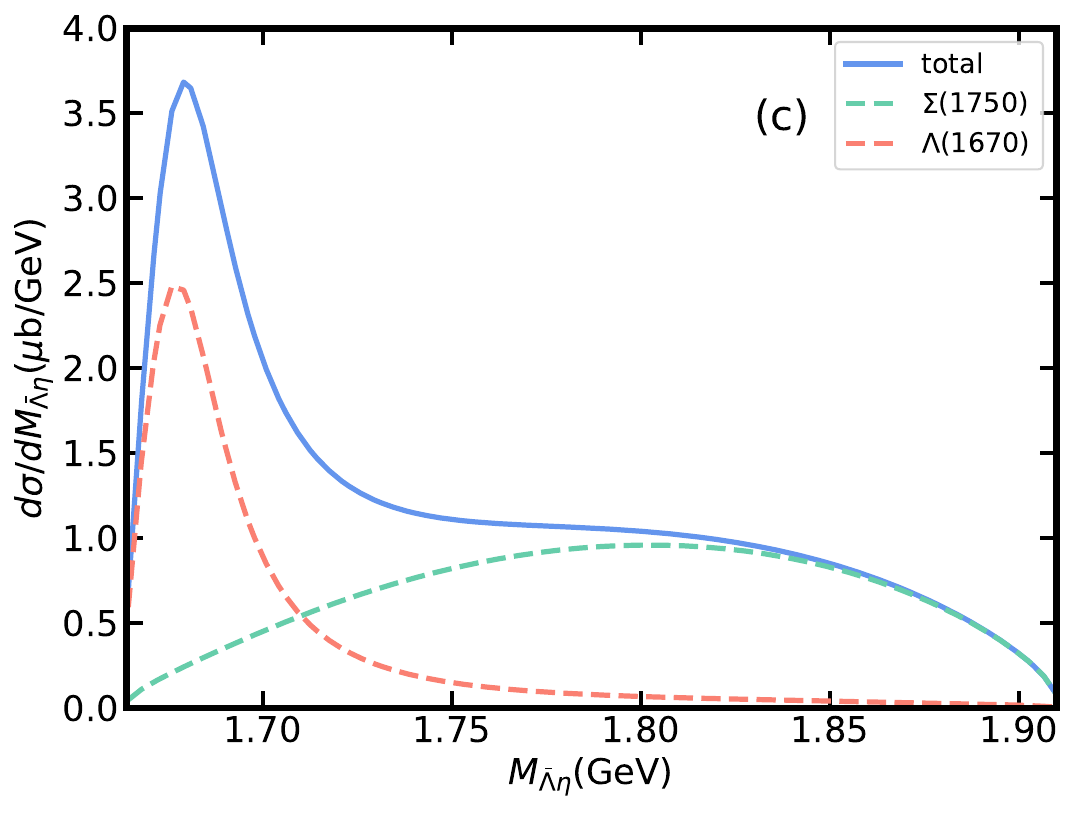}\hypertarget{6c}{}
	\includegraphics[width=0.3\textwidth]{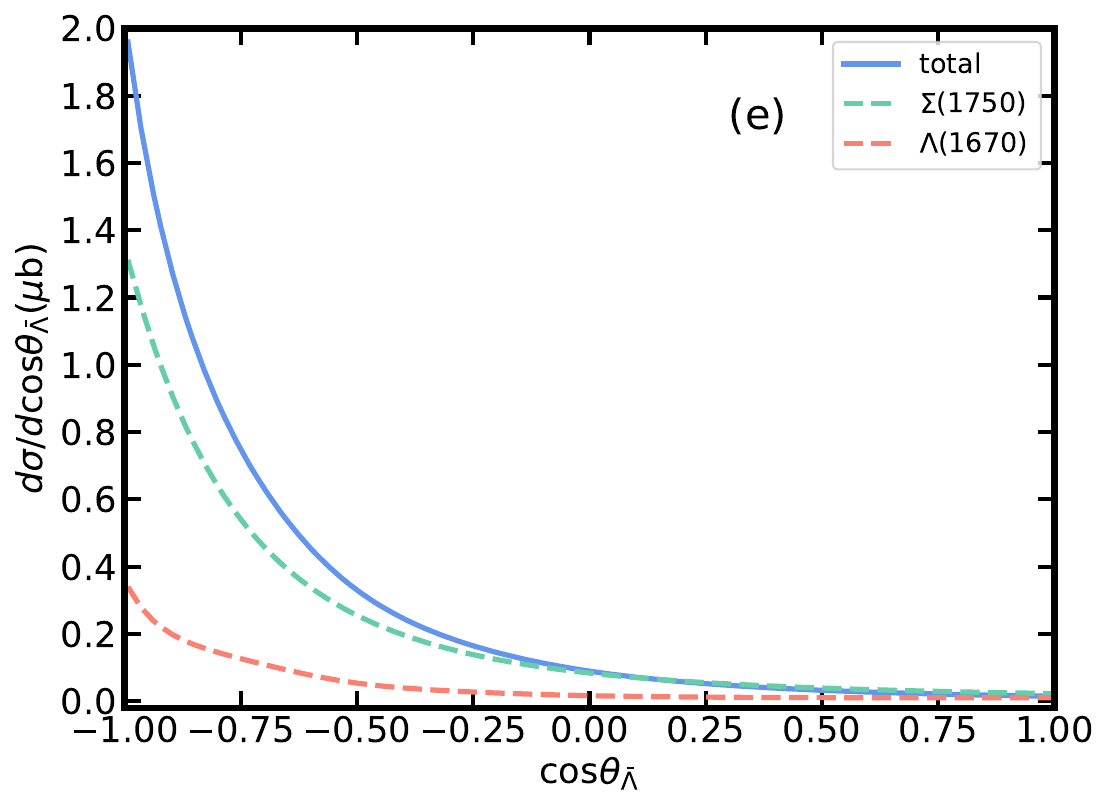}\hypertarget{6e}{}\\
	\includegraphics[width=0.3\textwidth]{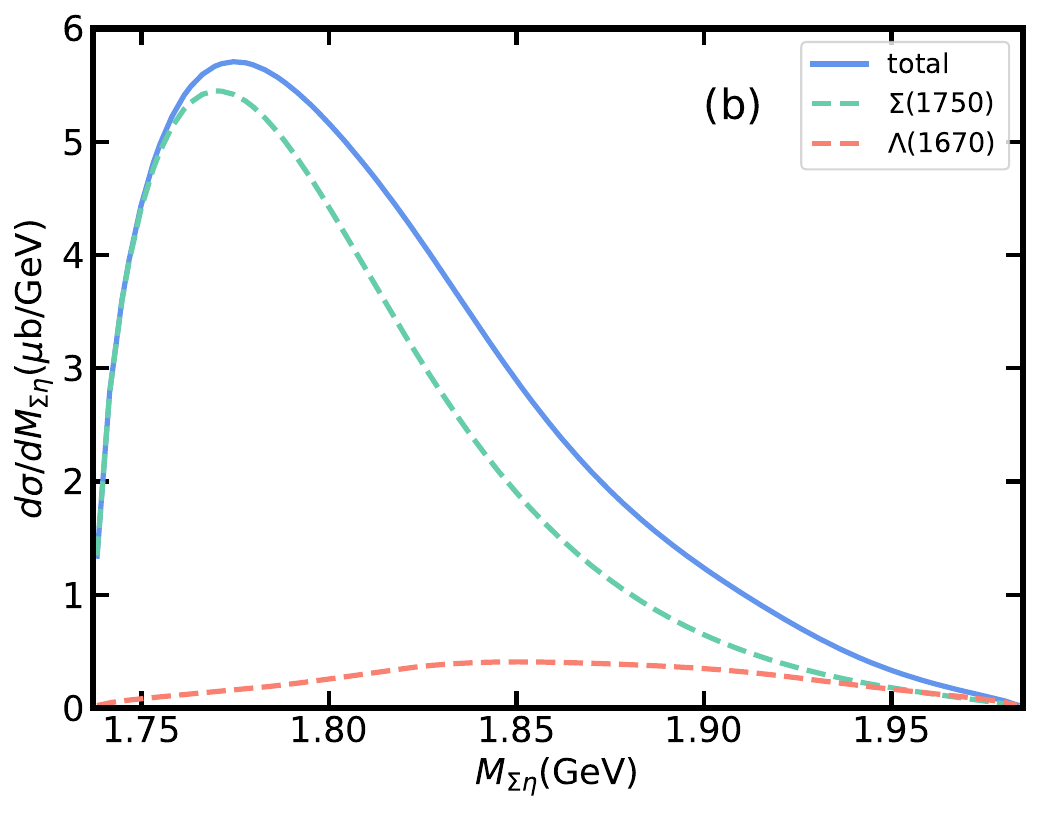}\hypertarget{6b}{}
	\includegraphics[width=0.3\textwidth]{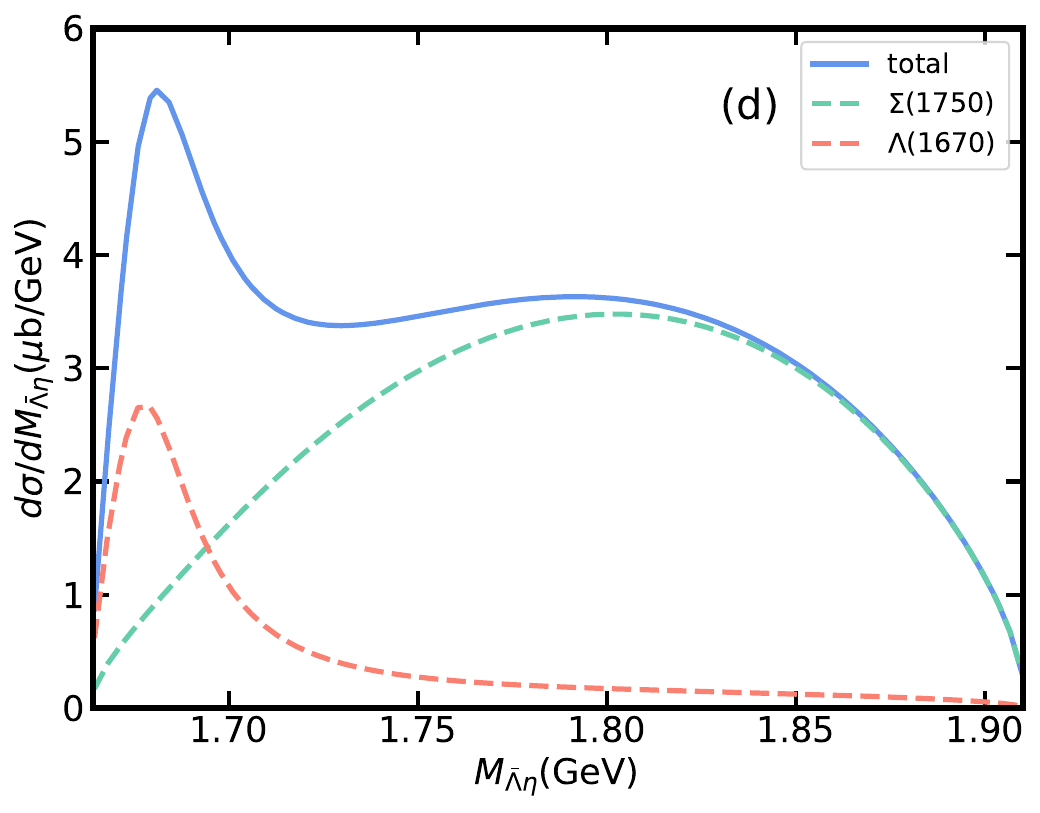}\hypertarget{6d}{}
	\includegraphics[width=0.3\textwidth]{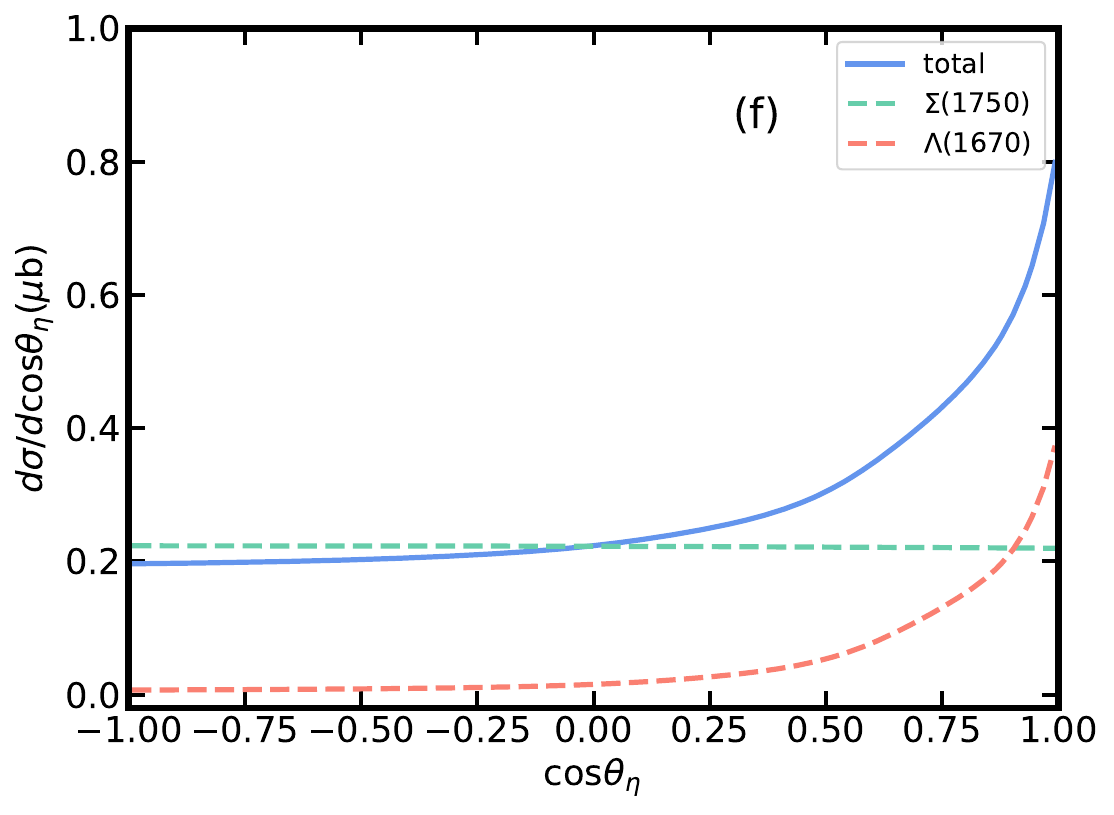}
	\captionsetup{justification=raggedright}
	\caption{At center of mass energy $W=3.1$~GeV, (a)-(b) invariant mass distribution of final $\Sigma\eta$ pair with $Br(\Sigma^*\to\Sigma\eta)=15\%$ and $55\%$, respectively; (c)-(d) invariant mass distribution of final $\bar{\Lambda}\eta$ pair with $Br(\Sigma^*\to\Sigma\eta)=15\%$ and $55\%$, respectively; (e) the angular distribution of $\bar{\Lambda}$; (f) the angular distribution of $\eta$ in the $\Sigma\eta$ rest frame.}
	\label{6}
\end{figure*}
In addition to the total cross section, we also study the differential cross section of $p\bar{p}\to\bar{\Lambda}\Sigma\eta$ with the center of mass energy $W=3.1$~GeV, shown in Fig.~\ref{6}, where $\Sigma(1750)$ is the main contribution. As can be seen from Fig.~\hyperlink{6a}{6(a)} and \hyperlink{6b}{6(b)}, there is an obvious peak of $\Sigma(1750)$ contribution in $\Sigma\eta$ invariant mass distribution. However, due to the influence of $\Lambda(1670)$, the peak energy of the total contribution is slightly higher than that in $\Sigma(1750)$. If we adopt $Br(\Sigma^*\to\Sigma\eta)=15\%$, $\Lambda(1670)$ will enhance the total contribution more significantly. The change in the coupling constant due to the decay width makes the total contribution for $Br(\Sigma^*\to\Sigma\eta)=55\%$ about 4 times that for $Br(\Sigma^*\to\Sigma\eta)=15\%$. Fig.~\hyperlink{6c}{6(c)} and \hyperlink{6d}{6(d)} are $\bar{\Lambda}\eta$ invariant mass distributions, from which the peak of $\Lambda(1670)$ can be clearly seen, and the branching ratio of $\Sigma(1750)\to\Sigma\eta$ does not affect the visibility of $\Lambda(1670)$ peak. This shows that even if $\Lambda(1670)$ contribution in total cross section is smaller than $\Sigma(1750)$, it is still possible to study its role by studying the invariant mass distribution of the $\bar{\Lambda}\eta$ system. In Fig.~\hyperlink{6e}{6(e)}, the dominant role of $\Sigma(1750)$ in the $u-$channel is clearly shown as the backward enhancement of the angular distribution of $\bar{\Lambda}$. Compared to $\Sigma(1750)$, the contribution of $\Lambda(1670)$ is small in both the $\bar{\Lambda}$ and $\eta$ angular distributions, but its forward angle of the $\eta$ angular distribution significantly affects the shape of the total contribution.

\begin{figure*}[htbp]
	\centering
	\includegraphics[width=0.3\textwidth]{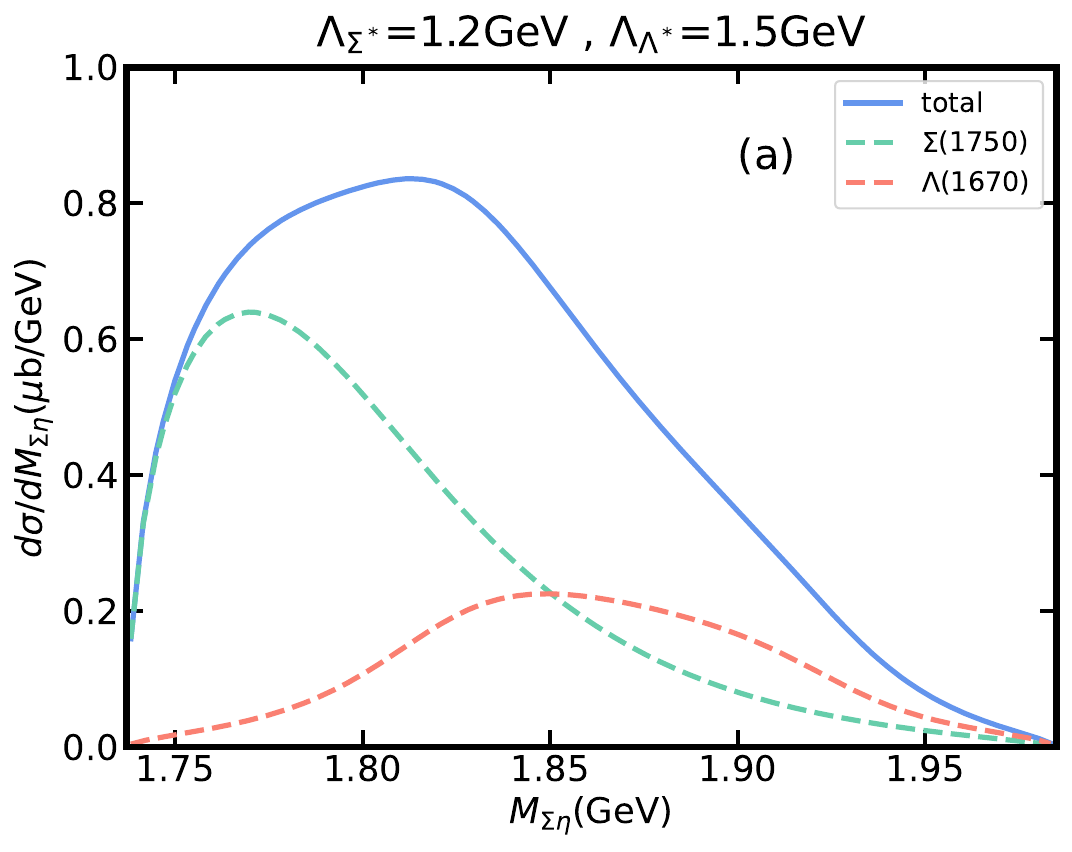}\hypertarget{7a}{}
	\includegraphics[width=0.3\textwidth]{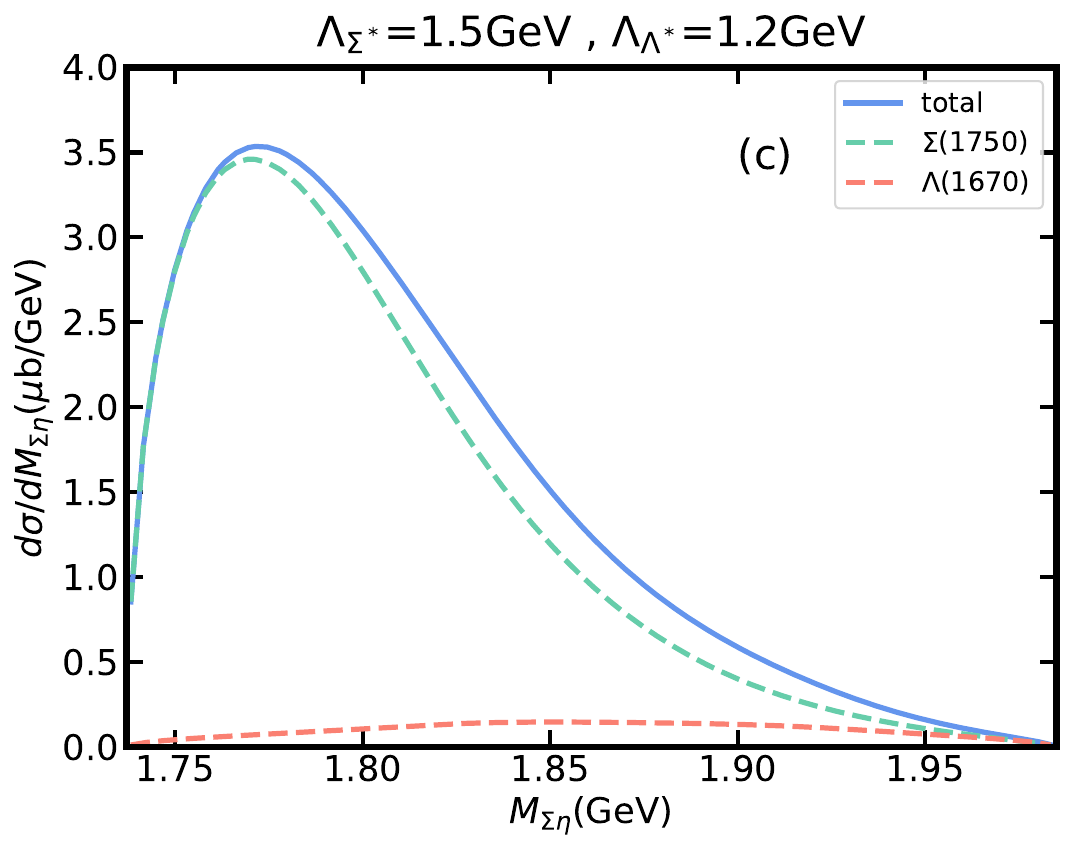}
	\includegraphics[width=0.3\textwidth]{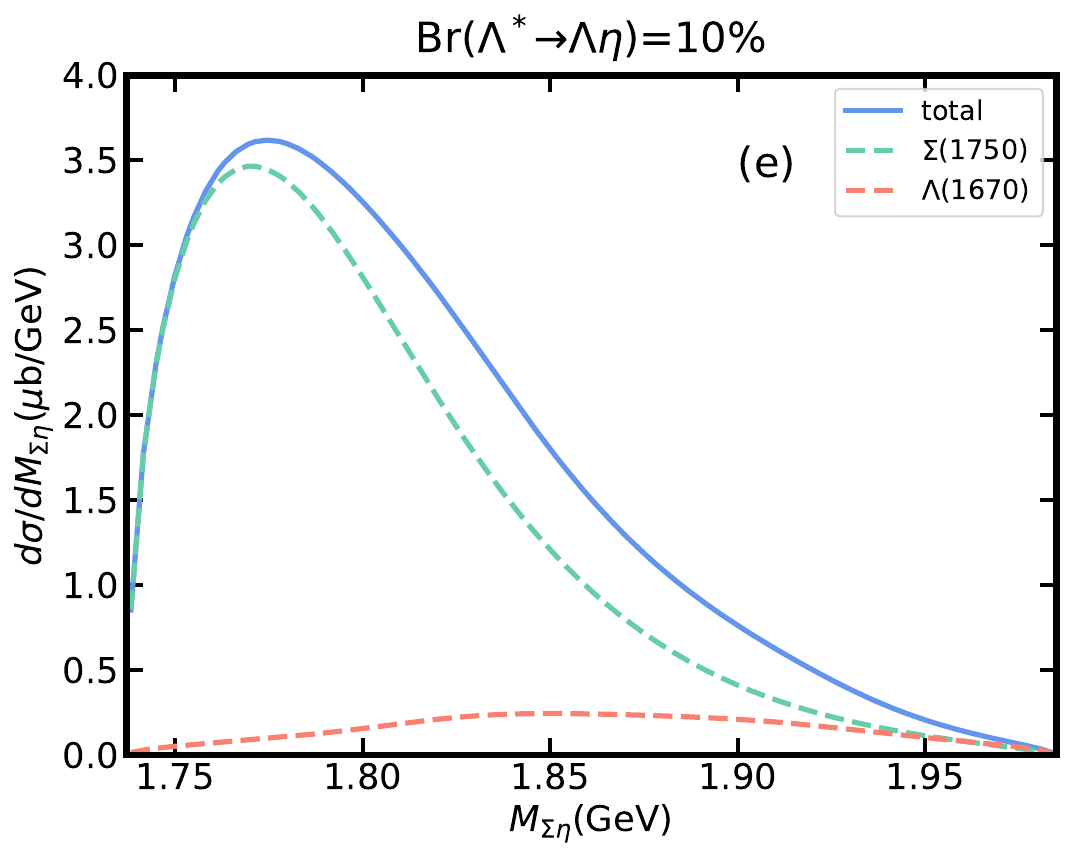}\hypertarget{7e}{}\\
	\includegraphics[width=0.3\textwidth]{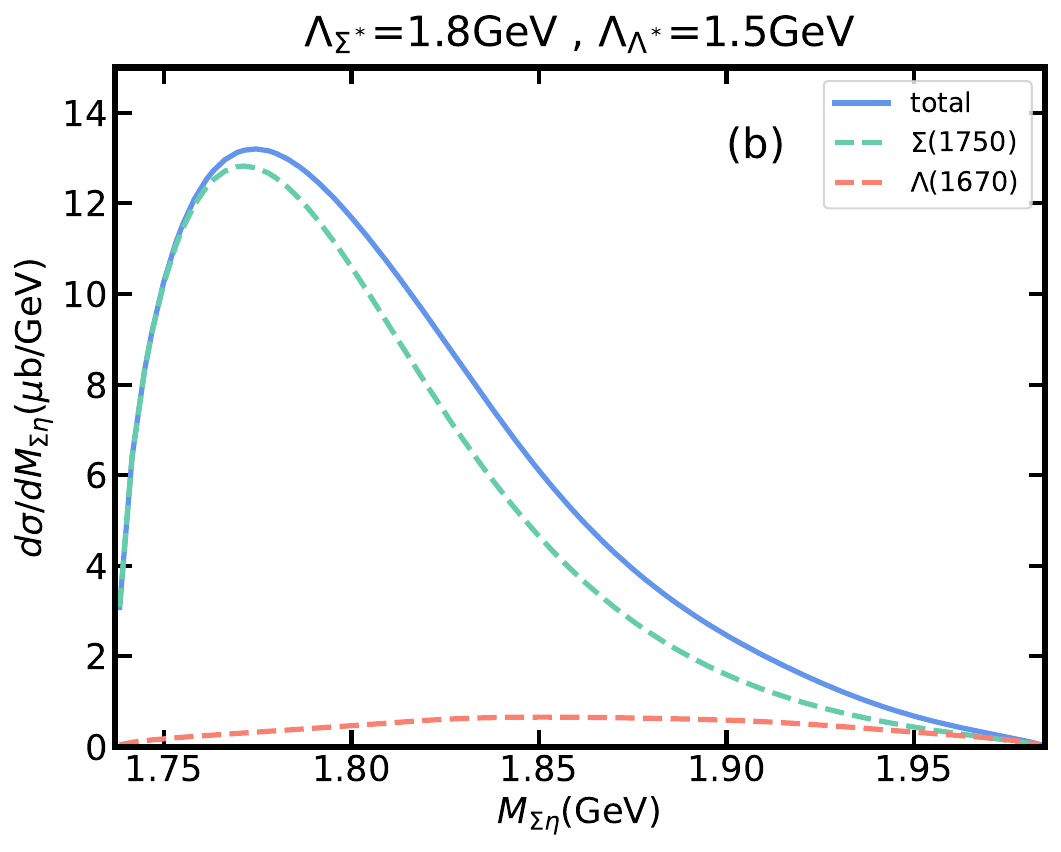}
	\includegraphics[width=0.3\textwidth]{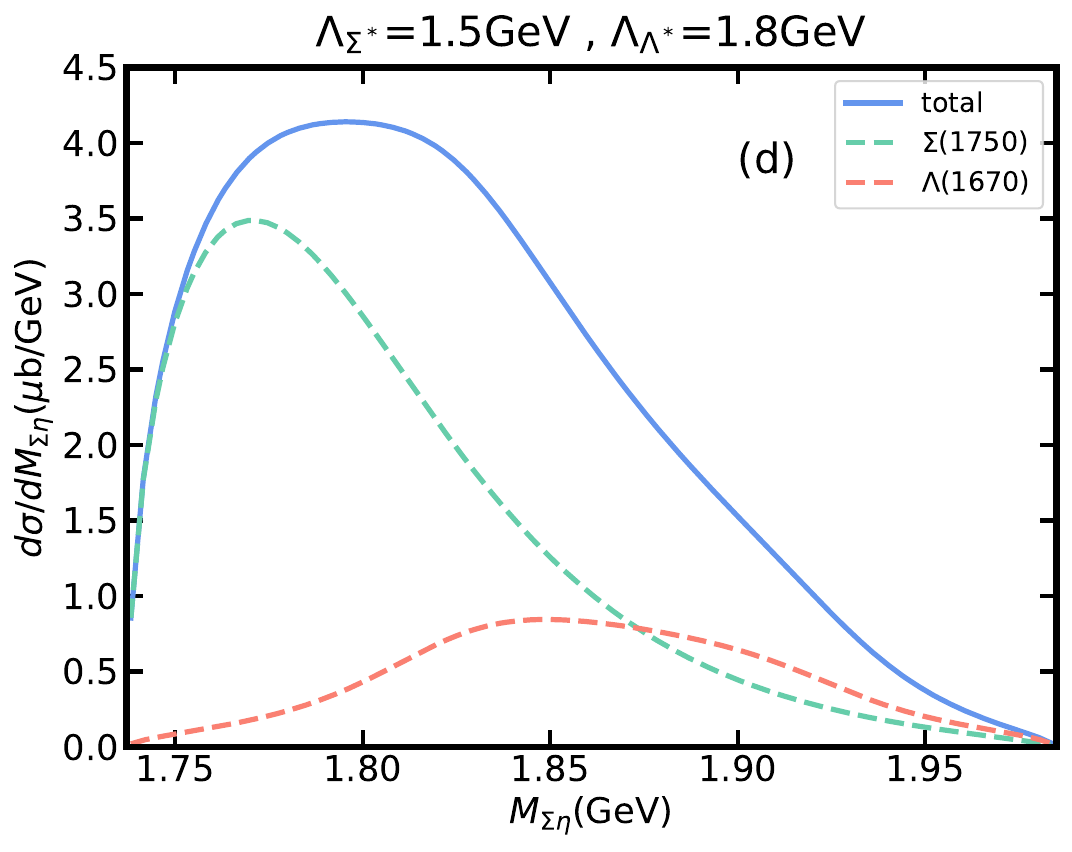}
	\includegraphics[width=0.3\textwidth]{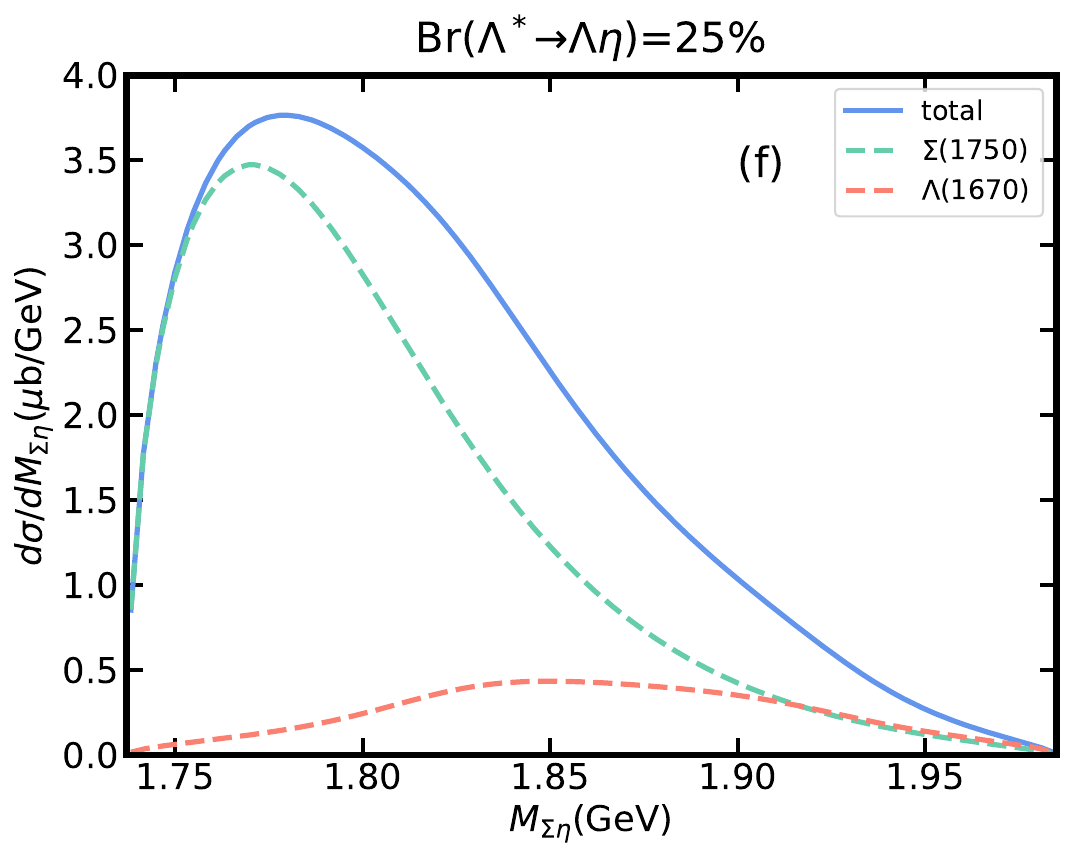}
	\captionsetup{justification=raggedright}
	\caption{At center of mass energy $W=3.1$~GeV, the invariant mass distributions of final $\Sigma\eta$ pair for $p\bar{p}\to\bar{\Lambda}\Sigma\eta$ reaction under different conditions, where (a)-(b) focus on the effect of $\Lambda_{\Sigma^*}$; (c)-(d) focus on the effect of $\Lambda_{\Lambda^*}$; (e)-(f) focus on the effect of $Br(\Lambda^*\to\Lambda\eta)$.}
	\label{7}
\end{figure*}
\begin{figure}[htbp]
	\centering
	\includegraphics[width=0.45\textwidth]{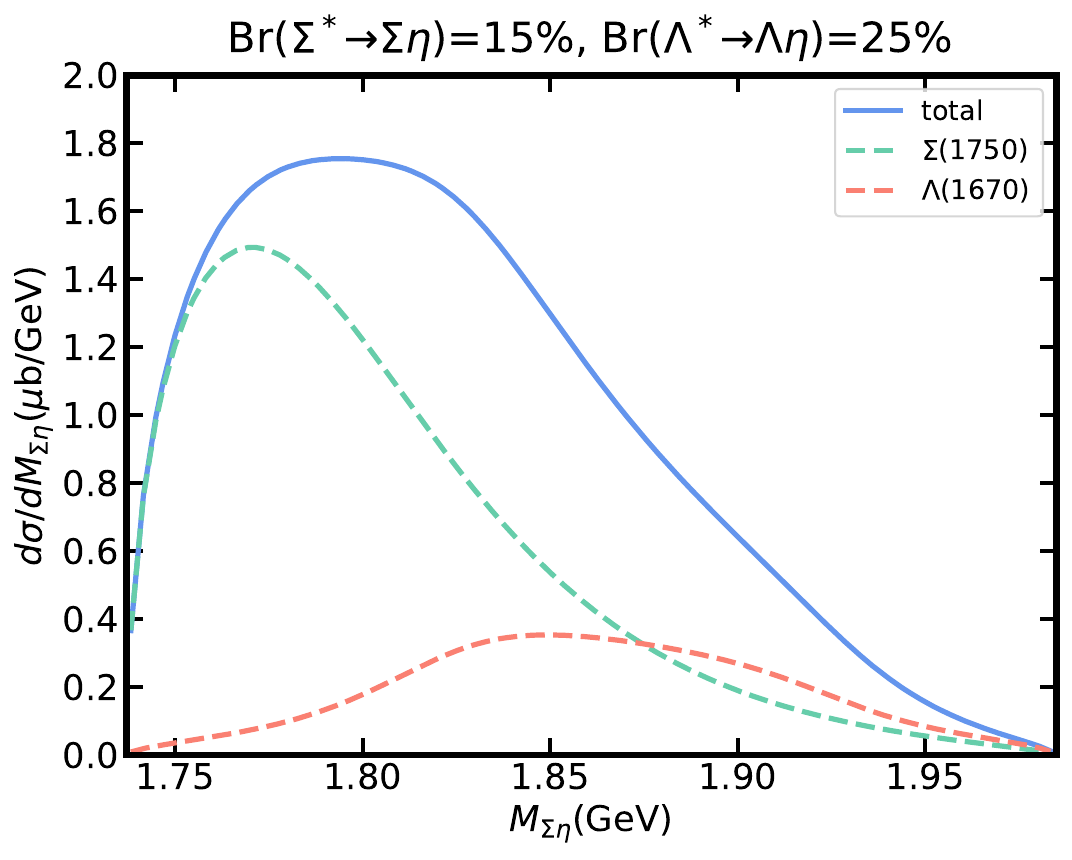}
	\captionsetup{justification=raggedright}
	\caption{At center of mass energy $W=3.1$~GeV, the invariant mass distribution of final $\Sigma\eta$ pair in the worst case with $Br(\Sigma^*\to\Sigma\eta)=$15\%, $Br(\Lambda^*\to\Lambda\eta)=$25\%.}
	\label{8}
\end{figure}
Similarly, the dependence of the peak position in $\Sigma\eta$ invariant mass distribution on the model parameters, $\Lambda_{\Sigma^*}$ and $\Lambda_{\Lambda^*}$, also need to be considered. In Fig.~\hyperlink{7a}{7(a)-(d)}, we show the results of $\Sigma\eta$ invariant mass distribution by fixing the cut-off parameter $\Lambda_{\Sigma^*}=1.5$~GeV or $\Lambda_{\Lambda^*}=1.5$~GeV and making another cut-off parameter vary in 1.2, 1.8~GeV. It is obvious that the peak position of total contribution in $\Sigma\eta$ invariant mass distribution depends on the selection of the cut-off parameters $\Lambda_{\Sigma^*}$ and $\Lambda_{\Lambda^*}$. However, the dominant role of the $\Sigma(1750)$ resonance in $\Sigma\eta$ invariant mass distribution will not be affected. Roughly, when $\Lambda_{\Sigma^*}>\Lambda_{\Lambda^*}$, the peak position of the total contribution will be closer to the result that $\Sigma(1750)$ resonance contributes alone. When $\Lambda_{\Sigma^*}<\Lambda_{\Lambda^*}$, the peak position of the total contribution will be significantly affected by $\Lambda(1670)$. Especially when $\Lambda_{\Sigma^*}=1.2$~GeV, $\Lambda_{\Lambda^*}=1.5$~GeV, i.e. Fig.~\hyperlink{7a}{7(a)}, the energy of peak position of the total contribution will increase significantly, and $\Sigma(1750)$ will form a shoulder-shaped structure to the left of the total contribution peak. The effect of the branching ratio for $\Lambda(1670)\to\Lambda\eta$ decay is shown in Fig.~\hyperlink{7e}{7(e)-(f)}, where $\Lambda(1670)$ contribution slightly affects the peak position and width of the total contribution. Moreover, we also predict the results of $\Sigma\eta$ invariant mass distribution in the worst case, i.e. $Br(\Sigma^*\to\Sigma\eta)=$15\%, $Br(\Lambda^*\to\Lambda\eta)=$25\%, shown in Fig.~\ref{8}. Even so, the total contribution is still completely dominated by the $\Sigma(1750)$ resonance, but the peak position and width are greatly affected by $\Lambda(1670)$.

\begin{figure}[htbp]
	\centering
	\includegraphics[width=0.45\textwidth]{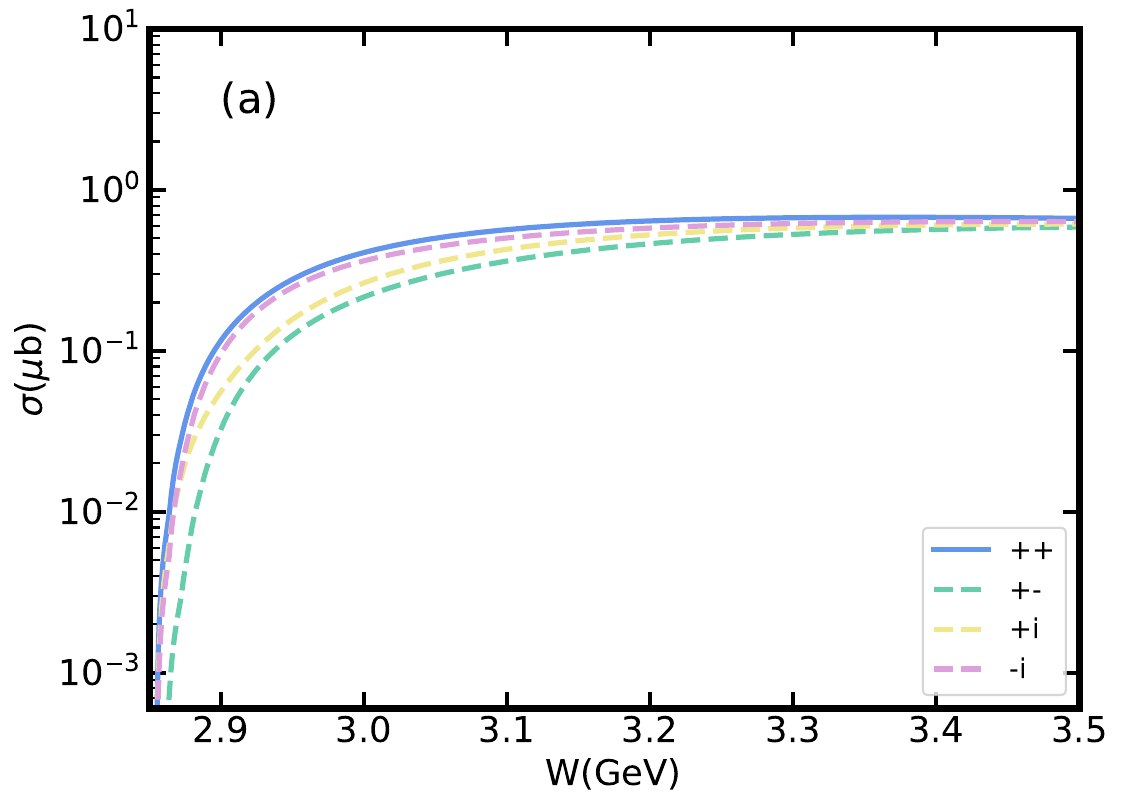}\hypertarget{9a}{}\\
	\includegraphics[width=0.45\textwidth]{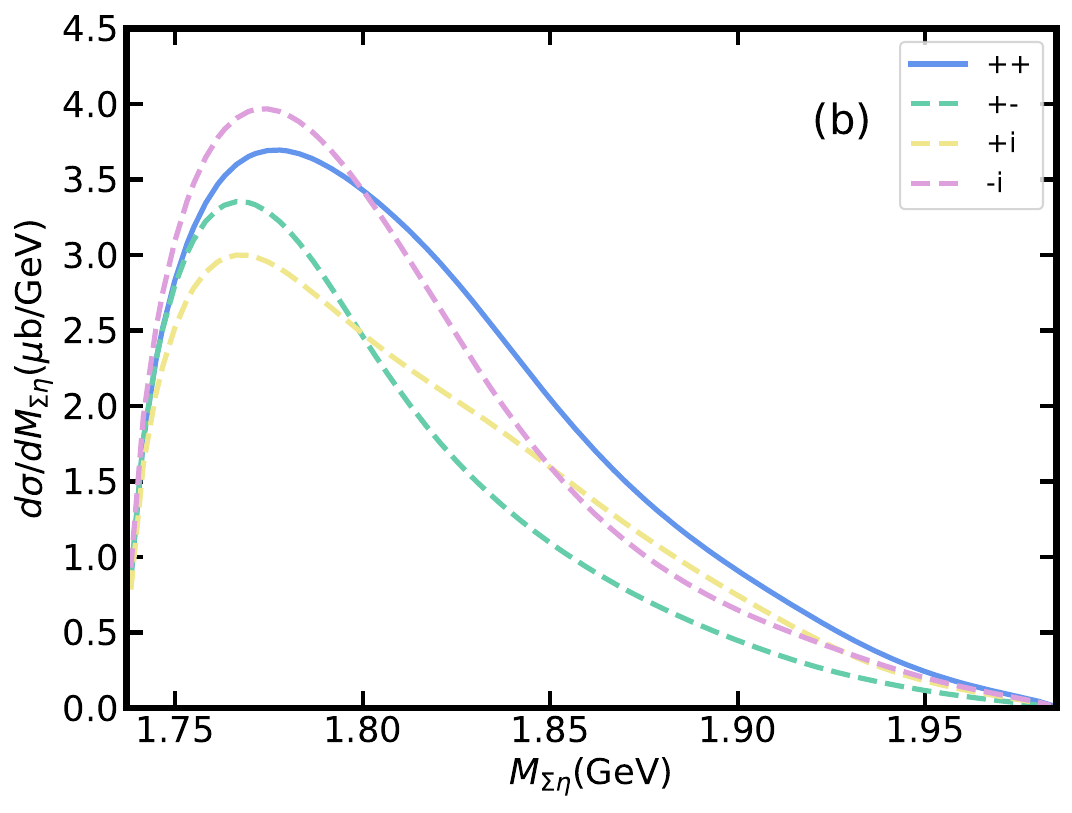}\hypertarget{9b}{}
	\captionsetup{justification=raggedright}
	\caption{The results of (a) total cross sections and (b) $\Sigma\eta$ invariant mass distributions at center of mass energy $W=3.1$~GeV with different factors introduced.}
	\label{9}
\end{figure}
Finally, we shall discuss the possible interference effects on the results. In our model, we extract the coupling constants of the vertex associated with the $\Sigma(1750)$ from the decay width. However, this approach makes it impossible to determine the relative phase between $\Sigma(1750)$ and $\Lambda(1670)$. Therewith, we consider to detect the existence of interference effects by multiplying factors of -1, i, -i. The results present in Fig.~\ref{9} where "++" represents not multiplying by any factor; "+-", "+i", "-i" represent $\Lambda(1670)$ contribution multiplied by factors of -1, i, -i, respectively. As can be seen from Fig.~\hyperlink{9a}{9(a)}, the influence of interference effect on the total cross section is not significant, it will not affect the dominant role of $\Sigma(1750)$. However, the interference effect will change the shape, peak position and width of the $\Sigma\eta$ invariant mass distribution. Especially when $\Lambda(1670)$ multiplied by the factor of i, the right side of the peak will show a very gentle sloping structure.

\section{conclusion}
We have made a theoretical study of $p\bar{p}\to\bar{\Lambda}\Sigma\eta$ reation baesd on an effective Lagrangian approach. In our model, we consider the production of $\Sigma(1750)$ and $\Lambda(1670)$ as intermediate states excited by the $K$ and $K^*$ meson exchanges between the initial proton and antiproton. We provide a prediction of total and differential cross sections and discuss the possible influence of $\Sigma(1750)\Sigma\eta$ vertex coupling and model parameters. According to our results, $\Sigma(1750)$ resonance makes a significant contribution near the threshold, making the reaction suitable for studying the features of $\Sigma(1750)$ resonance. The cut-off parameters have significant effect on the position and width of the peak in $\Sigma\eta$ invariant mass distribution, but $\Lambda_{\Sigma^*}$ and $\Lambda_{\Lambda^*}$ are usually regarded as free parameters, therefore, more experimental data are needed to determine them. The branching ratio for $\Lambda(1670)\to\Lambda\eta$ decay has a slight effect on $\Sigma\eta$ invariant mass distribution. Moreover, the interference effect can also significantly affect the shape of $\Sigma\eta$ invariant mass distribution and the position and width of the peak.

\section{acknowledgments}
H.S. is supported by the National Natural Science Foundation of China (Grant No.12075043). X.L. is supported by the National Natural Science Foundation of China under Grant No.12205002.

\bibliographystyle{unsrt}
\bibliography{b}
\end{document}